\documentclass[conference]{IEEEtran}
\IEEEoverridecommandlockouts
\usepackage{cite}
\usepackage{amsmath,amssymb,amsfonts}
\usepackage{algorithmic}
\usepackage{graphicx}
\usepackage{textcomp}
\usepackage{xcolor}
\usepackage{booktabs}
\usepackage{makecell}
\usepackage{listings}
\usepackage{xcolor} 
\usepackage{xurl}
\usepackage[breaklinks=true]{hyperref} 
\usepackage{placeins}
\def\BibTeX{{\rm B\kern-.05em{\sc i\kern-.025em b}\kern-.08em
    T\kern-.1667em\lower.7ex\hbox{E}\kern-.125emX}}
\lstdefinestyle{IEEE_code_style}{
    basicstyle=\ttfamily\small,  
    commentstyle=\color{gray},
    keywordstyle=\bfseries,       
    stringstyle=\color{purple},
    breaklines=true,              
    frame=tb,                     
    showstringspaces=false,
    captionpos=b                  
}

\begin{document}

\title{Retrieval Quality at Context Limit\\
}

\author{\IEEEauthorblockN{Max McKinnon}
\IEEEauthorblockA{\textit{Google LLC} \\
Mountain View, U.S.A. \\
mckinnonm@google.com}
}

\maketitle

\begin{abstract}
The ability of large language models (LLMs) to recall and retrieve information from long contexts is critical for many real-world applications. Prior work (Liu et al., 2023) reported that LLMs suffer significant drops in retrieval accuracy for facts placed in the middle of large contexts, an effect known as ``Lost in the Middle'' (LITM). We find the model Gemini 2.5 Flash can answer needle-in-a-haystack questions with great accuracy regardless of document position including when the document is nearly at the input context limit. Our results suggest that the ``Lost in the Middle'' effect is not present for simple factoid Q\&A in Gemini 2.5 Flash, indicating substantial improvements in long-context retrieval.
\end{abstract}



\bstctlcite{IEEEexample:BSTcontrol}  

\section{Introduction}
Large language models (LLMs) have rapidly advanced in their ability to process and reason over long textual contexts, enabling applications in summarization, retrieval-augmented Q\&A, document understanding, and more. However, earlier work \cite{liu2023lost} highlighted a limitation: when provided with very large contexts, models like GPT-3.5 and Llama-2 exhibited significant drops in retrieval accuracy for facts positioned in the middle of the context window when approaching their context limits, a phenomenon dubbed ``Lost in the Middle'' (LITM). This effect is attributed to inherent primacy and recency biases caused by architectural constraints such as causal attention masks and relative positional encodings \cite{mit2025unpacking, chen2023extending, yang2024found}. While a larger context window theoretically enables better information access, it also increases the risk of critical details being obfuscated or ignored \cite{munn2024context}.

Recent models such as Gemini 2.5 Flash, Claude Sonnet 4, and GPT 4.1 offer context windows of 1M+ tokens. Recent literature suggests that although needle-in-a-haystack performance has improved, more complex forms of failure remain such as in multimodal \cite{vaidya2025multimodal} or multi-needle \cite{langchain2024multi}.

In this work, we analyze the 1M+ token model Gemini 2.5 Flash for retrieval performance on single-factoid retrieval (needle-in-a-haystack) questions. Contrary to prior LITM results that showed recall performance correlated with position of the factoid, we observe great recall for Gemini 2.5 Flash: all 26 facts are retrieved correctly, regardless of position.

\section{Methods}
We use the transcript of Friends \cite{kaggle:friends-script} as the starting corpus exceeding the input context limit of Gemini 2.5 Flash: it is 924k words long and exceeds 1M Gemini tokens. The context limit is approximately 700k English words from the estimate of 60-80 English words per 100 Gemini tokens \cite{Google_Tokens_2025_misc}.

We create 20 unique non-canon dialogue snippets to insert throughout the transcript, such as Monica sharing to Phoebe her favorite ice cream flavor is honey lavender. From these 20 unique non-canon factoids, 26 unique questions and answers are made against them. Transcript length is controlled by trimming the end off of the full transcript such that the context after adding the snippets back in reaches the desired size. For each trimmed transcript, the snippets are evenly distributed from beginning to end of the corpus with one at the beginning and one the end.

For each injected snippet of information, we write one or two question-answer pairs targeting that information, e.g. ``What is Monica's favorite ice cream flavor?'' In total, we present 26 questions.

Each question is tested for other correct answers when the injected information was ablated, for example asking without the non-canon honey lavender ice cream conversation injected resulted in this response which shows that the information only came from that single needle-in-a-haystack injection of information: 
``Although Monica was occasionally depicted enjoying ice cream, such as in flashback scenes of her as Fat Monica, these instances did not specify a particular flavor she favored above all others. Her culinary expertise and discerning palate were often showcased through the elaborate meals and desserts she prepared for her friends, but her personal ice cream preference remains an unconfirmed detail for even the most devoted viewers.''

The Gemini 2.5 Flash API returns a 400 error when an input request exceeds the maximum token limit, and that behavior is used to precisely validate the size of the context limit. We test retrieval in steps of 10\% of the full Friends transcript, and we find 70\% is able to run but 80\% was not. At 80\%, the error message is ``The input token count (1105498) exceeds the maximum number of tokens allowed (1048576)''. 70\% of the full transcript 924k words, 647k words, aligns with the estimate of 700k derived from the average word per token estimate.

In this study, we test Gemini 2.5 Flash. We also considered Claude Sonnet 4 and GPT 4.1, however the lower priced tiers artificially constrain the input context limit, and obtaining the higher token input tiers of both to get to their true 1M token context limit is outside the budget of this study. As a workaround, we attempted to run the experiment with the chat versions of each, but those both appear to have their own context limits after some testing.

We submit all 26 questions simultaneously with the context file's text in a single text query to a fresh instance of a model. Temperature is set to 0.1, thinking is disabled, and no system instructions are given. Each answer is saved and compared to ground truth with a fresh instance of Gemini 2.5 Flash and manually verified for accuracy. 

\section{Results}

\begin{table}[htbp]
\centering
\caption{Out of 26 asked questions, Gemini 2.5 Flash gets all 26 questions correct all the way up to its context limit in steps of 10\% of the full transcript. The context limit was exceeded at 80\% of the full transcript.}
\label{tab:llm_comparison}
\begin{tabular}{cccc}
\toprule
\textbf{\makecell{Ratio of \\ Full Transcript}}  & \textbf{\makecell{Ratio of \\ Context Limit}} & \textbf{\makecell{Questions \\ Correct}} & \textbf{Accuracy} \\
\midrule
0.1 & 0.13 & 26 & 1.0 \\
0.2 & 0.26 & 26 & 1.0 \\
0.3 & 0.40 & 26 & 1.0 \\
0.4 & 0.53 & 26 & 1.0 \\
0.5 & 0.66 & 26 & 1.0 \\
0.6 & 0.79 & 26 & 1.0 \\
0.7 & 0.92 & 26 & 1.0 \\
0.8 & 1.05 & N/A & N/A \\

\bottomrule
\end{tabular}
\end{table}

\section{Discussion}
Our results indicate that the LITM phenomenon, as observed in 2023 models, appears to have been substantially mitigated or eliminated in Gemini 2.5 Flash for direct Q\&A over injected information. Although it is not the same dataset under test, Gemini 2.5 Flash had perfect accuracy at Q\&A based upon single factoid (needle-in-a-haystack) retrieval at all context sizes tested within the 1M token context limit.

We hypothesize that the lack of an LITM result stems from positional encoding improvements such as Attention with Linear Biases (ALiBi) \cite{press2022trainshorttestlong}, training curriculum changes stressing needle-in-a-haystack performance \cite{geminiteam2024gemini15unlockingmultimodal}, and potentially other sources not yet identified.

\subsection{Literature Comparison}
Comparing to foundational studies, (Liu et al., 2023) established the U-shaped performance curve, showing maximum retrieval accuracy at the context’s start and end, with substantial degradation in the middle \cite{liu2023lost}.

Architecture bias plays a role. These effects are attributed to transformer architecture's causal attention and relative position encodings, especially Rotary Position Embeddings (RoPE), which causes long-distance decay and middle-context loss \cite{mit2025unpacking, chen2023extending, yang2024found}

Recent advances have changed the behavior of Q\&A retrieval. Google's Gemini 1.5 and 2.5 models show near-perfect accuracy on single needle-in-a-haystack retrieval tasks, even at maximum context length \cite{bastian2024gemini}, \cite{geminiteam2024gemini15unlockingmultimodal}, \cite{comanici2025gemini25pushingfrontier}.

Needle-in-a-haystack is only one type of retrieval and new benchmarks caution that single-needle results may not generalize to harder benchmarks. New benchmarks focus on multi-needle, reasoning, and modality crossing \cite{vaidya2025multimodal, langchain2024multi, chen2025helmet}.

Modern best practices can prioritize critical information in a much smaller context such as Retrieval-Augmented Generation (RAG), but the performance falloff can return in multi-needle and multimodal tasks \cite{langchain2024multi, promptingguide2025rag, olamendy2023lost}.

\subsection{Limitations}
There are several limitations of these results. Only factoid Q\&A is tested; no paraphrased or ambiguous queries are tested. The context is only text, not audio or mixed modalities. The Q\&A supporting factoids were unique unrelated facts; competing and conflicting facts were not included.

\subsection{Future Work}
We identify several areas of future work: test with paraphrased, ambiguous, or adversarial queries; add multiple contradictory or distractor facts; explore multi-hop and reasoning tasks, not just lookup Q\&A; repeat with audio or mixed modality input; explore the parameter space on retrieval performance such as model temperature; and explore smaller models in the context of a system solution such as Retrieval-Augmented Generation (RAG) involving a search retrieval step and a smaller language model answering the questions.

\section{Conclusion}
We find no evidence of a context position effect for simple Q\&A fact retrieval in modern Gemini 2.5 Flash even up to input sizes just under its million-token scale. For practical single-needle Q\&A tasks over long documents, state-of-the-art now appears to exhibit near-perfect recall. Further research should focus on solving the more subtle forms the problem has evolved into such as multi-needle reasoning and multimodal tasks.

\section{Acknowledgements}
I would like to thank Elliot Patros for feedback on this work and code reviews. I would also like to thank Meenakshi Barjatia, Nikhil Bhanu, and Akshay Cadambi for technical discussion and motivation related to this work.

\appendices
\section{Injected Information}
\label{app:snippets}
Twenty snippets of non-canon factoids were injected in order and equidistantly across the Friends transcript. The transcript was about half the size of the full Friends transcript to fill the input context limit. Here is an example of one such injected factoid:

\begin{quotation}
\setlength{\parindent}{0pt}  
\setlength{\parskip}{1ex}   
\itshape  
Guy on the street: Hey Joey, did you ever learn to play the banjo?
Joey: Not yet, but I signed up for lessons last week!
\end{quotation}

\section{Example Question with Ground Truth Answer}
\label{app:yaml-data}

The ground-truth question-answer pairs corresponding to the injected 
snippets were structured in YAML, as shown in Listing \ref{lst:yaml}.

\begin{lstlisting}[caption={Ground-Truth Q\&A Pairs.}, label={lst:yaml}]
- question: What instrument did Joey sign up to learn?
  answer: Banjo.
\end{lstlisting}

\bibliography{references}

\end{document}